\documentclass{mem}
\usepackage{natbib}
\usepackage{txfonts}
\usepackage{balance}
\usepackage{graphicx}
\usepackage{flushend}
\usepackage[a4paper,breaklinks,pdftex]{hyperref}
\idline{75}{282}
\DeclareUnicodeCharacter{2215}{ }
\begin{document}
\def\teff{$T\rm_{eff }$}
\def\kms{$\mathrm {km s}^{-1}$}

\title{
CUBES: a UV spectrograph for the future
}
\subtitle{}

\author{
S.\,Covino\inst{1} 
\and
S.\,Cristiani\inst{2,3} 
\and
J. M. Alcalá\inst{4}
\and
S. H. P. Alencar\inst{5}
\and
S. A. Balashev\inst{6}
\and
B. Barbuy\inst{7}
\and
N. Bastian\inst{8,9}
\and
U. Battino\inst{10}
\and
L. Bissell\inst{11}
\and
P. Bristow\inst{12}
\and
A. Calcines\inst{13}
\and
G. Calderone\inst{2}
\and
P. Cambianica\inst{14}
\and
R. Carini\inst{15}
\and
B. Carter\inst{16}
\and
S. Cassisi\inst{17,18}
\and
B. V. Castilho\inst{19}
\and
G. Cescutti\inst{20}
\and
N. Christlieb\inst{21}
\and
R. Cirami\inst{2}
\and
R. Conzelmann\inst{12}
\and
I. Coretti\inst{2}
\and
R. Cooke\inst{22}
\and
G. Cremonese\inst{14}
\and
K. Cunha\inst{23,24}
\and
G. Cupani\inst{2,3}
\and
A.R.\,da Silva\inst{25}
\and
D. D'Auria\inst{4}
\and
V. De Caprio\inst{4}
\and
A. De Cia\inst{26}
\and
H. Dekker\inst{27}
\and
V. D’Elia\inst{28}
\and
G. De Silva\inst{29}
\and
M. Diaz\inst{7}
\and
P. Di Marcantonio\inst{2}
\and
V. D'Odorico\inst{2,3}
\and
H. Ernandes\inst{7}
\and
C. Evans\inst{30}
\and
A. Fitzsimmons\inst{31}
\and
M. Franchini\inst{2}
\and
B. G\"ansicke\inst{32}
\and
M. Genoni\inst{1}
\and
R.~E.~Giribaldi\inst{25}
\and
C. Gneiding\inst{19}
\and
A. Grazian\inst{14}
\and
C. J. Hansen\inst{33}
\and
J. Hopgood\inst{12}
\and
J. Kosmalski\inst{12} 
\and
F. La Forgia\inst{34}
\and
P. La Penna\inst{12}
\and
M. Landoni\inst{1}
\and
M. Lazzarin\inst{34}
\and
D. Lunney\inst{11}
\and
W. Maciel\inst{7}
\and
W. Marcolino\inst{35}
\and
M.Marconi\inst{4}
\and
A. Migliorini\inst{36}
\and
C. Miller\inst{11}
\and
A. Modigliani\inst{12} 
\and
P. Noterdaeme\inst{37,38}
\and
L. Oggioni\inst{1}
\and
C. Opitom\inst{39}
\and
G. Pariani\inst{1}
\and
B.~Pilecki\inst{25}
\and
S. Piranomonte\inst{15}
\and
A. Quirrenbach\inst{21}
\and
E.M.A. Redaelli\inst{1}
\and
C. B. Pereira\inst{23}
\and
S. Randich\inst{40}
\and
S. Rossi\inst{7}
\and
R. Sanchez-Janssen\inst{11}
\and 
M. Schoeller\inst{12}
\and
W. Seifert\inst{21}
\and
R. Smiljanic\inst{25}
\and
C. Snodgrass\inst{39}
\and
O. Squalli\inst{12}
\and
I. Stilz\inst{21}
\and
J. St\"urmer\inst{21}
\and
A. Trost\inst{2,20}
\and
E. Vanzella\inst{41}
\and
P. Ventura\inst{15}
\and
O. Verducci\inst{19}
\and
C. Waring\inst{11}
\and
S. Watson\inst{11}
\and
M. Wells\inst{11}
\and
D. Wright\inst{16}
\and
T. Zafar\inst{29}
\and
A. Zanutta\inst{1}
\and
G. Zins\inst{12}
}

\institute{
INAF - Osservatorio Astronomico di Brera, via E. Bianchi 46, 23807 Merate, Italy --
\email{stefano.covino@inaf.it}
\and
INAF - Osservatorio Astronomico di Trieste, via G. B. Tiepolo 11, 34131 Trieste, Italy
\and
IFPU – Institute for Fundamental Physics of the Universe, Via Beirut 2, 34151 Trieste, Italy
\and
INAF-Osservatorio Astronomico di Capodimonte, via Moiariello 16, 80131 Napoli, Italy
\and
Departamento de Fisica - ICEx - UFMG, Belo Horizonte, MG, Brazil
\\
The remaining affiliations can be found at the end of the paper.
}

\authorrunning{Covino et al.}

\titlerunning{The CUBES spectrograph}

\date{Received: Day Month Year; Accepted: Day Month Year}

\abstract{
In spite of the advent of extremely large telescopes in the UV/optical/NIR range, the current generation of 8-10m facilities is likely to remain competitive at ground-UV wavelengths for the foreseeable future. 
The Cassegrain U-Band Efficient Spectrograph (CUBES) has been designed to provide high-efficiency ($>40$\%)  observations in the near UV (305-400 nm requirement, 300-420 nm goal) at a spectral resolving power of $R>20,000$, although a lower-resolution, sky-limited mode of $R \sim 7,000$ is also planned.

CUBES will offer new possibilities in many fields of astrophysics, providing access to key lines of stellar spectra: a tremendous diversity of iron-peak and heavy elements, lighter elements (in particular Beryllium) and light-element molecules (CO, CN, OH), as well as Balmer lines and the Balmer jump (particularly important for young stellar objects). The UV range is also critical in extragalactic studies: the circumgalactic medium of distant galaxies, the contribution of different types of sources to the cosmic UV background, the measurement of H$_2$ and primordial Deuterium in a regime of relatively transparent intergalactic medium, and follow-up of explosive transients.

The CUBES project completed a Phase A conceptual design in June 2021 and has now entered the Phase B dedicated to detailed design and construction. First science operations are planned for 2028. In this paper, we briefly describe the CUBES project development and goals, the main science cases, the instrument design and the project organization and management.

}
\maketitle{}

\section{Introduction}
\label{sec:intro}  
The four 8.2m telescopes of the Very Large Telescope (VLT) at the European Southern Observatory (ESO)
form the world’s most scientifically productive ground-based observatory in the visible and infrared. However, looking at the future of the VLT, there is a long-standing need for an optimised ultraviolet (UV) spectrograph \citep{Barbuy2014} with a large increase of efficiency with respect to existing instruments (UVES and X-Shooter).

\begin{figure*}[t!]
\resizebox{\hsize}{!}{\includegraphics[clip=true]{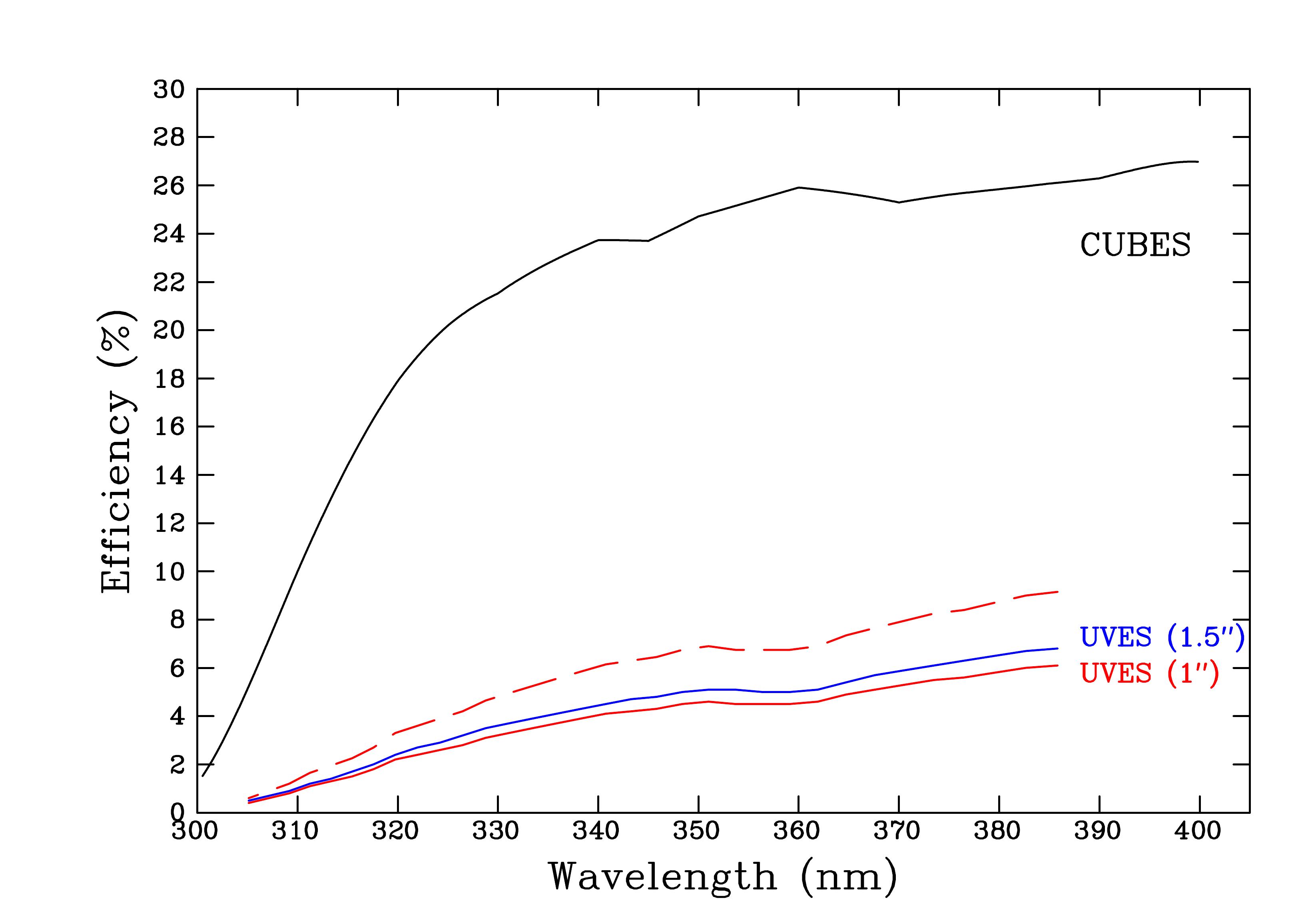}}
\caption{\footnotesize
{\bf Comparison of the total (instrument+telescope+sky) CUBES efficiency in the $R>20$\,K mode with respect to UVES.}}
\label{fig:cubelt}
\end{figure*}

The European Extremely Large Telescope (ELT), under construction in northern Chile by ESO, with a primary aperture of 39m will be unprecedented in its light-gathering power, coupled with exquisite angular resolution via correction for atmospheric turbulence by adaptive optics (AO). At variance with current large telescopes such as the VLT, AO is an integral part of the ELT, which has a five-mirror design including a large adaptive mirror (M4) and a fast tip-tilt mirror (M5). The choice of protected silver (Ag+Al) for the ELT mirror coatings (apart from M4) ensures a durable, proven surface with excellent performance across a wide wavelength range. However, the performance drops significantly in the blue-UV part of the spectrum compared to bare aluminium. ESO is actively researching alternative coatings, but in the short-medium term we can assume that the performance of the ELT in the blue-UV will be limited. Indeed, during the Phase A study of the MOSAIC multi-object spectrograph \citep{Evans2016} it was concluded that a blue-optimised instrument on the VLT could potentially be competitive with the ELT at wavelengths shorter than 400 nm (Fig.\,\ref{fig:cubelt}). In addition, this spectral range is complementary to the ELT and JWST. Motivated by this, in 2018 we revisited \citep{Evans2018} the Phase A study undertaken in 2012 of the Cassegrain U-band Brazilian-ESO Spectrograph. The past study investigated a $R\sim 20$k spectrograph operating at ‘ground UV’ wavelengths (spanning 300-400 nm) to open-up exciting new scientific opportunities compared to the (then) planned instrumentation suite for Paranal Observatory \citep{Barbuy2014,Bristow2014}. 

In January 2020 ESO issued a Call for Proposal for a Phase A study of a UV Spectrograph to be installed at a Cassegrain focus of the VLT, with the goal of high-efficiency ($>40$\%) and intermediate resolving power ($\sim$ 20K) in the ground-UV domain (305-400 nm requirement, 300-420 nm goal). 
In May 2020 the CUBES (Cassegrain U-Band Efficient Spectrograph) Consortium, led by INAF\footnote{Istituto Nazionale di AstroFisica}, was selected to carry out the study. 
The CUBES project completed a Phase A conceptual design study in June 2021.
After the endorsement by the ESO Council at the end of 2021,
Phase B started in February 2022 with the signature of the Construction Agreement between ESO and the leading institute of the CUBES Consortium, opening the detailed design and construction phase. Here we report the present status of the project, which will provide a world-leading UV capability for ESO from 2028 well into the ELT era. More detailed information about the project is reported in \citet{Cristiani2022SPIE}.

\section{Science with CUBES}
\label{sec:science} 
The CUBES science case covers a wide range of astrophysical topics. We propose below a brief highlight of the main key cases across the Solar System, Galactic, and extra-galactic fields \citep[see also ][]{Evansetal2022}.

\subsection{Searching for water in the asteroid belt}
The search for water in our solar system is a long-standing problem \citep{Opitom2022}. It is a difficult task with ground-based facilities, given the water content of Earth's atmosphere. The typical diagnostics of water outgassing from small bodies is the OH emission at 308 nm. Observation of the OH line has been possible so far for a few active comets while they are near the Sun and Earth, with severe limitations. We still miss knowledge of water production around their orbits and the role of seasonal effects that the Rosetta mission revealed to be important. In general, most comets are simply too faint to be studied with current facilities. Main-belt comets, bodies in asteroidal orbits, can undergo activity thought to arise from sublimation. Constraining the OH emission of these objects is well beyond our current capabilities. Since main-belt comets show a size distribution similar to the general population in the asteroid belt, the detection of outgassing water with CUBES would point to a potentially large population of icy bodies. This could imply a large reservoir of water, a parameter of considerable interest in models of the formation and evolution of the inner solar system.

\subsection{Accretion, winds \& outflows in YSOs}
The evolution of circumstellar disks, mass accretion around young stars, and outflow and winds are fundamental aspects of the formation of protoplanets. Observations about these phenomena require multi-wavelengths studies of stars during the first 10\,Myr of their evolution and in particular of Classical T Tauri stars (CTTS). These young, low- to solar-mass stars are actively accreting mass from planet-forming disks. Spectroscopic surveys of CTTS in nearby star-forming regions have been carried out to study the often complex relationships between accretion, jets and disk structure. CUBES, both due to its increased UV sensitivity and coverage of a critical wavelength range, will enable more detailed studies of the accretion processes/wind-outflows than currently possible as well as studies of CCTS in star-forming region at larger distances.

\subsection{Bulk composition of exo-planets} 
In the past few decades, we have learned that it is normal for stars to have planets and the study of exoplanet formation and evolution has become a major astrophysical topic. The best approach available at present for estimating the bulk composition of exo-planet systems is based on spectroscopic analysis of evolved white dwarf (WD) stars accreting debris from tidally-disrupted planetesimals. WDs are hot so most of their abundance diagnostics are in the near-UV (e.g. Sc, Ti, V, Cr, Mn, Fe, Ni). However, WDs are also intrinsically faint, and only about twenty systems have precise abundances so far. CUBES will transform this exciting area of exo-planet research by increasing the sample of known exo-planetesimal compositions providing precise constraints on the next generation of planet-formation models.

\subsection{Stellar nucleosynthesis}
\begin{figure*}[t!]
\resizebox{6.9cm}{!}{\includegraphics[clip=true]{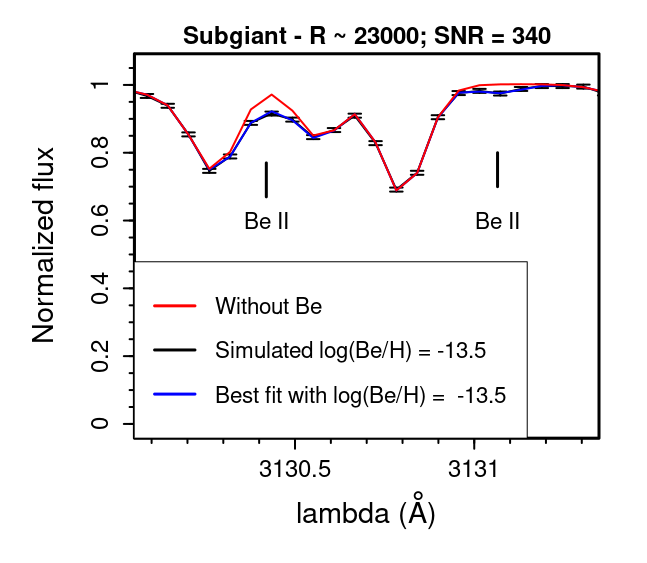}}  \resizebox{6.9cm}{!}{\includegraphics[clip=true]{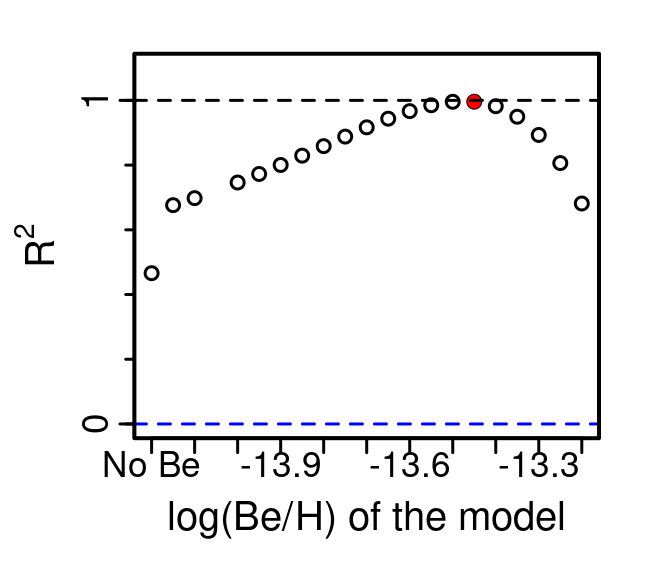}} \\
\center{ \resizebox{7.4cm}{!}{\includegraphics[clip=true]{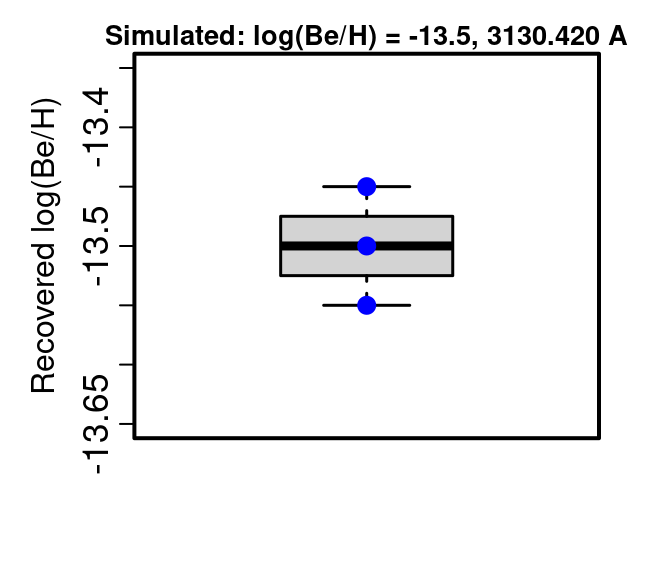}}}
\caption{\footnotesize
{\bf Results of fitting four simulated observations of a bright (V = 12.5\,mag) subgiant star (T$_{\rm eff}$ = 5600\,K and $\log$ g = 3.4 dex) with [Fe/H] = -3.5 and $\log$(Be∕H) = -13.5. The final simulated observations have
different realizations of SNR = 340 and were computed to have R = 23\,K
and sampling of 2.35\,px. 
{\it Top Left:} One of the best fit synthetic spectra is shown in the plot and has the same Be abundance of the simulated observation. {\it Top Right:} The best fit is chosen using the coefficient of determination, $R^2$, which involves the ratio between the residual sum of squares and the total sum of squares. {\it Bottom:} The boxplot displays the best fitting Be abundances for each one of the four simulated observations (with results deviating at most by 0.05 dex from the input Be abundance).}}
\label{fig:cubesbe}
\end{figure*}

The spectral features of more than a quarter of the chemical elements are only observable in the near UV, but the low efficiency of instruments in this domain severely restricted previous studies. Advancements in the field require high-resolution, near-UV spectroscopy of a large number and diversity of stars. Three main CUBES science cases deal with this topic:

i) {\it Metal-poor stars and light elements}. A key case is to probe the early chemical evolution of the Galaxy, via chemical abundance patterns in the oldest, low-mass stars that formed from material enriched by the first supernovae. The so-called Carbon-enhanced metal-poor (CEMP) stars are the perfect probes to investigate nucleosynthesis by the first stars. CUBES will enable quantitative spectroscopy for large samples of metal-poor stars, providing direct estimates for a broad range of heavy elements, as well as valuable constraints on CNO elements. 

ii) {\it Heavy-element nucleosynthesis}. Stellar abundances from CUBES will provide critical tests of the various production channels of heavy elements for both r- and s-process elements. Determining the abundances of neutron-capture elements in metal-poor stars is fundamental to understand the physics of these processes and the chemical evolution of the Galaxy as well as the origin of the Galactic halo. Since lines due to many of these elements are in the UV domain (e.g. Hf, Sn, Ag, Bi) and have only been measured for a very restricted number of stars, CUBES will play a critical role to fill this gap.

iii) {\it Beryllium is one of the lightest and simplest elements}. Nevertheless, questions remain about its production in the early Universe. Recent results are consistent with no primordial production, but larger samples are required to investigate this further. Only $\sim 200$ stars have Be abundances so far \citep[limited to $V$ $\sim$ 12 mag in a few hrs with UVES, ][]{UVES2000}. CUBES will provide large homogeneous samples of Be abundances in stars belonging to different populations up to three magnitudes deeper, providing new insights into its production and tracing the star-formation history of the Galaxy (see Fig.\,\ref{fig:cubesbe}). 

\subsection{Primordial deuterium abundance}
Within the Standard Model of particle physics and cosmology there is still no accepted model for dark energy and dark matter, or why the Universe contains baryons instead of antibaryons, or even why the Universe contains baryons at all. We are also missing crucial properties of neutrinos (e.g. their hierarchy, why they change flavour, and the number that existed during the earliest phases of the Universe). 
Some of these questions can be investigated by measuring the nuclides produced a few minutes after the Big Bang. The primordial deuterium (D/H) abundance is currently our most reliable probe of Big Bang Nucleosynthesis \citep{Cooketal2014}. CUBES will provide a large, reliable sample of D/H estimates from quasar absorption spectra. Its significant gain at the shortest wavelengths compared to existing facilities will enable observations at lower redshifts (less contamination by the Lyman-alpha forest) giving more absorption-line systems from which to estimate D/H and smaller uncertainties.

\subsection{The missing baryonic mass at the cosmic noon}

Remarkable progresses about the missing baryon problem has been recently made possible at low redshifts by studying the dispersion measure in Fast Radio Bursts (FRBs) \citep{Macquart2020} and at $z > 1.5$ by observations and simulations of the Lyman forest \citep[e.g., ][]{Weinberg1997}. Still, we have insufficient knowledge about how baryonic matter is distributed among the different gaseous components and we would need to better constrain the mechanisms (stellar and AGN feedback, accretion, etc.) that determine the observed distribution. A UV efficient spectrograph with relatively high resolution offers the possibility to dig into the complex nature of the inter- and circum-galactic gas at $z \sim$ 1.5 to 3, via two experiments with quasar absorption lines:
i) The baryons in the diffuse IGM are studied through the detection and analysis of Lyman-$\alpha$ lines at $z \simeq 1.5$ to 2.3. This redshift range, immediately after the era of peak star-formation in the Universe, is poorly investigated due to observational difficulties as the low efficiency of ground-based spectrographs in the UV, but is critical to connect the low- and high-redshift results. 
ii) Observing the O VI absorption lines at $1.9 < z < 2.9$ to trace the warm-hot gas at $T > 10^5$ K, associated with the IGM or with the CGM \citep{Lehner2014}. In all cases spectroscopy in the wavelength range at $\lambda > 400$ nm is necessary to complement the information on the neutral IGM component derived from the
Lyman forest, checking the associated metal absorption lines (in particular due to C IV and Si IV) and deriving the contribution of the ionised gas. It is also needed to complete the coverage of the associated HI and metal transitions. To this aim spectra of the same targets obtained at higher resolution with, e.g., UVES/VLT (simultaneously via a fiber link or retrieved from the archives) will be needed.

\subsection{Cosmic UV background}
Galaxies are likely able to produce most of the UV emissivity needed for cosmic reionisation at high redshift but quasars also possibly contribute. Estimates of the escape fraction (f$_{\rm esc}$) of hydrogen-ionising photons able to escape a galaxy are close to 100\% for quasars. However, the volume density of low- and intermediate-luminosity quasars at $z$ $>$ 4 is still uncertain, so it is unclear if they are the dominant source of ionisation. In contrast, star-forming galaxies are more numerous, but estimates of f$_{\rm esc}$ from observations of the Lyman continuum ($\lambda_{\rm rest}$ $<$ 91.2 nm) have uncertainties of tens of percent and are limited to a handful of systems at $z$ = 2.5 to 4. To be detectable from Earth escaping photons have to survive absorption along the line of sight by the intergalactic medium, which become stronger with redshift and is significantly variable between sightlines. Given these competing factors, the ideal redshift range for ground-based observations of the Lyman continuum of a galaxy is $z$ = 2.4 to 3.5, i.e. from about 410\,nm down to the atmospheric cut-off. For this reason CUBES could be an asset for this science case thanks to its high throughput. Furthermore, since the galaxies to be observed are extremely faint this science case is also one of the main drivers of the low resolution mode.

\subsection{Transient astronomy}
Time-domain astronomy is one of the most active branches of modern astrophysics. In a few years, new observational facilities, specifically designed with the goal of securing high-cadence observations of large fractions of the nightly sky, will become operational. Equally important, ``big-data'' algorithms are increasingly being applied and developed to manage the large amount of data provided by these facilities. The discovery space opened by rare or peculiar transients is very large, involving all categories of sources. For low or high redshift objects, a highly efficient UV spectrograph can shed light on a variety of physical ingredients and, in this context, the possible synergy of the CUBES and UVES spectrographs could open the exciting perspective of a UV-blue continuous spectral coverage.

\section{From Science to Requirements}
\label{sec:requirements}  
The science cases of interest for the CUBES community defined the reference for the development of the Top level Requirements (TLR):
\begin{itemize}
\item Spectral range: CUBES must provide a spectrum of the target over the entire wavelength range of 305 – 400 nm in a single exposure (goal: 300 – 420 nm).
\item Eﬃciency: The eﬃciency of the spectrograph, from slit to detector (included), has to be $>40$\% for 305 – 360 nm (goal $>45$\%, with $>50$\% at 313 nm), and $>37$\% (goal $40$\%) between 360 and 400 nm.
\item Resolving power ($R$): In any part of the spectrum, $R$ needs to be $>19$\,K, with an average value $>20K$.
\item Signal-to-noise (S/N) ratio: In a 1 hr exposure the spectrograph must be able to obtain, for an A0-type star of $U$ = 17.5 mag (goal $U \ge 18$ mag), a S/N = 20 at 313\,nm for a 0.007\,nm wavelength pixel.
\end{itemize}

During Phase A studies an important addition was identified with the provision of a second (lower) resolving power (with $R \sim 7$k), to enable background-limited observations of faint sources where spectral resolution is less critical. 
There are also studies for a fiber link to UVES, in order to provide simultaneous observations at longer wavelengths. This would considerably broaden the scientific capabilities of the project.

\section{Instrument Design Overview}
\label{sec:design}  
\begin{figure*}[t!]
\resizebox{\hsize}{!}{\includegraphics[clip=true]{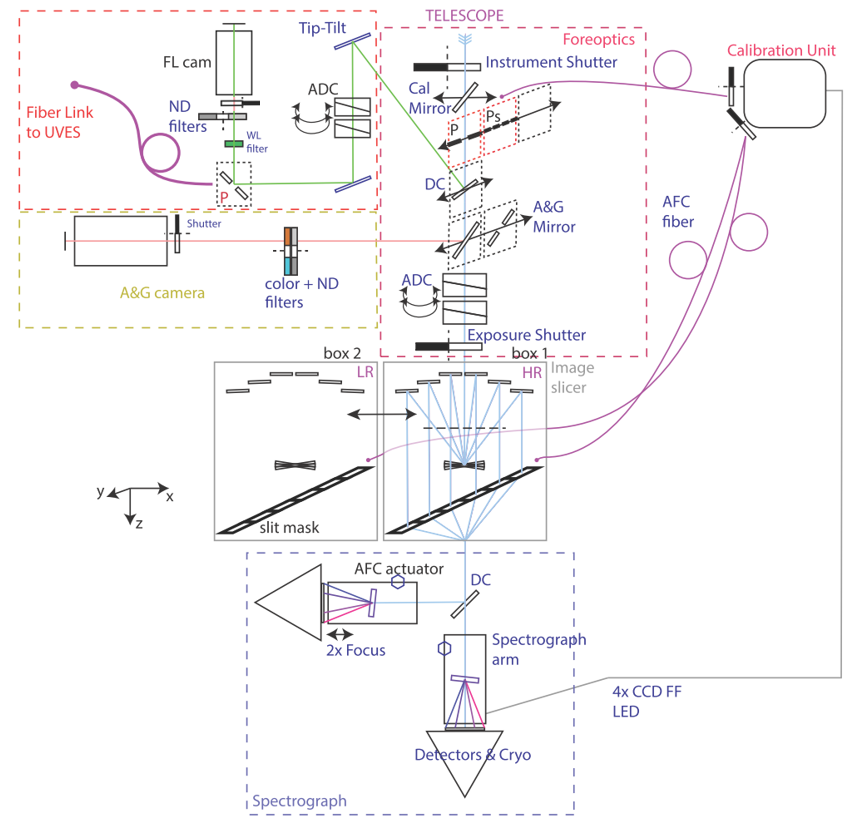}}
\caption{\footnotesize
{\bf Functional design of CUBES. Light path is from the top to the bottom. The shown acronyms are: FL cam (fibre-link A\&G camera), 
ND filters (Neutral Density filters), WL filter (Wavelength filter),
ADC (Atmospheric Dispersion Corrector), P  (single Pin-hole mask), Ps (multi pin-holes mask), A\&G (Acquisition and Guiding), Cal mirror (Calibration mirror), DC (Dichroic), AFC (Active Flexure Compensation),
LR (Low Resolution), HR (High Resolution), Cryo (Cryocooler),
FF LED (Flat-Field LED).}}
\label{fig:whcubes}
\end{figure*}

The instrument is designed to be used alone or combined with the fiber-feed to UVES. Two resolution modes (HR, $R > 20$\,K; LR $R \sim 7$\,K) are enable by the exchange of two independent image slicers. An Active Flexure Compensation system (AFC) is part of the baseline. All the top level requirements (TLRs), in particular those related to efficiency, are met (see Fig.\,\ref{fig:whcubes} for a view of the whole instrument).

\subsection{Instrument sub-systems and Operations}
The current baseline design of CUBES includes a
calibration unit that provides the light sources necessary to register frames for ﬂat ﬁelding, wavelength calibration, alignment, and the AFC. A foreoptics sub-system includes an atmospheric dispersion corrector (ADC) and Acquisition and Guiding functionalities. There are two user-selectable image slicers (to enable different spectral resolutions) followed by a dichroic beamsplitter that feeds two spectrograph arms, each consisting of a Fused Silica single-lens collimator, a first-order transmission grating with a groove density up to 3600 l/mm, and a 3-lens all-silica camera. Each arm has its own detector cryostat, with a 9k or 10k CCD detector (we also have an option to increase the sampling by 11\% using the STA 10k 9$\mu m$ detector instead of the E2V 9K 10$\mu m$), read-out electronics, cryo-vacuum components (both hardware and specific control electronics). The Instrument Control Electronics are based on PLCs that are compliant with  latest ELT design standard, and control all optomechanical and cryo/vacuum functions. The Scientific Detector system is controlled by ESOs new NGC2 standard. The Instrument Software package has elaborate instrument control software, data-reduction and simulation tools (see Sect.\ref{sec:Software}) for details). And, finally, a Fiber Link unit provides the option of simultaneous observations with UVES by injecting red light into optical fibers (1 object, 3 sky) that subtend 1 arcsec on the sky and are approximatively 40 m long to UVES on the Nasmyth platform \citep{UVES2000}.

\subsection{Optics and Dispersing Elements}
Using two lens doublets and a number of folding prisms the foreoptics relays a FoV of 6”x10” for LR and 1.5”x10” for HR at the telescope focus to the entrance plane of the spectrograph. Zero-deviation ADC prisms in the parallel beam between the doublets provide atmospheric dispersion correction over a range 300-405\,nm for zenith angles of 0-60$^\circ$. By inserting a dichroic just below the telescope focal plane, light redward of 420\,nm may be directed to the UVES fiber feed. 
At the magnified telescope focal plane produced by the fore-optics (0.5\,arcsec/mm plate scale), two user-selectable reflective image slicers are used to reformat the field of view. The first component, an array of six spherical slicer mirrors decomposes the rectangular FoV into six slices which are reimaged by six spherical camera mirrors composing the spectrograph entrance slit, defined by a slit mask. The slicer efficiency is expected to be $>90\%$ (goal 94\%) The output slit mask has six slitlets, corresponding to six slices, each one measuring 0.25”x10” on the sky for the HR slicer ($R \sim 20K$) and 1”x10” for the LR slicer ($R \sim 7K$). Further slit mask apertures are illuminated by a ThAr fiber source for use by the AFC system.

The light coming from the slit mask is folded by a Total Internal Reflection (TIR) prism and then reaches a dichroic which splits the light by reflecting the Blue-Arm passband (300–352.3\,nm) and transmitting the Red-Arm passband (346.3–405\,nm). 

In order to achieve a high ($>20K$) resolution without the efficiency losses associated with cross-dispersed echelles, CUBES uses state-of-the-art first-order dispersing elements. Binary transmission gratings produced by E-beam microlithography and an Atomic Layer Deposition (ALD) overcoat have been identified as a suitable technology \citep[see ][]{Zeitner2022}. Their theoretical average (RCWA) diffraction efficiency is $> 90\%$ and studies carried out through simulations and prototyping show that the measured efficiency is compliant with the expectation and demonstrate the feasibility of the instrument in terms of light throughput.

\subsection{Mechanics}
To achieve high resolution, CUBES requires a fairly large beam diameter of 160 mm and a collimator focal length of 3\,m.  In order to limit the total mass, a light-weight construction principle has been adopted. The optical layout contains several folds so all optical elements. The optical layout was optimized such that all optical elements of the spectrograph from slit to detector lie in a single plane, so all spectrograph optics can be mounted on a single optical bench. This is arguably the most stable configuration since the dispersion direction of  CUBES  is  parallel  to  the  stiff  surface  plane  of  the optical bench. A general focus of the mechanical design is, in fact, to minimize the effects of gravitational bending of the instrument.

\begin{figure*}[t!]
\resizebox{6.9cm}{!}{\includegraphics[clip=true]{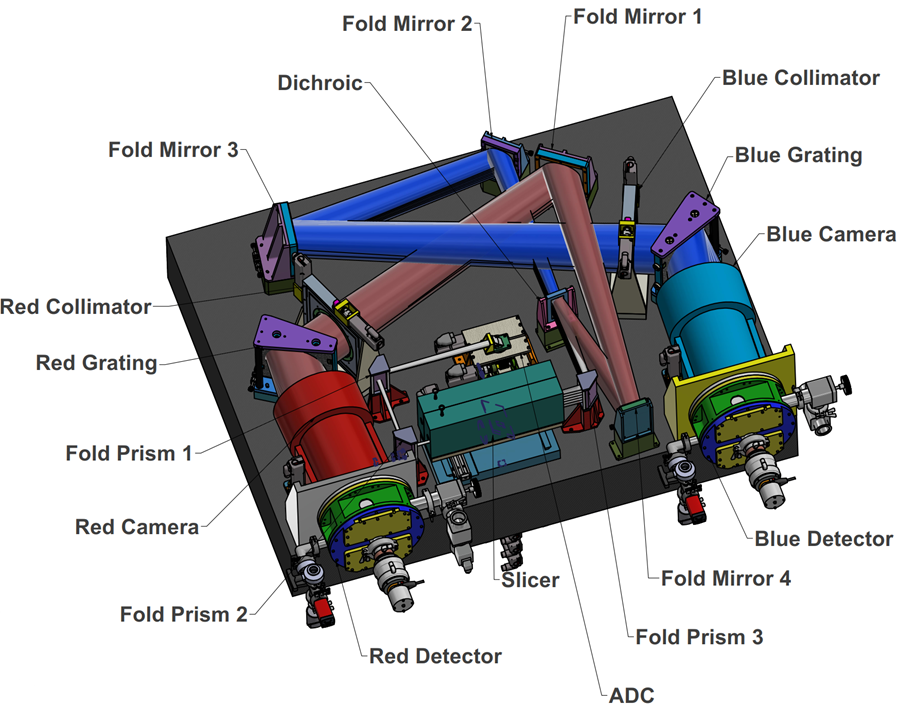}}  \resizebox{6.9cm}{!}{\includegraphics[clip=true]{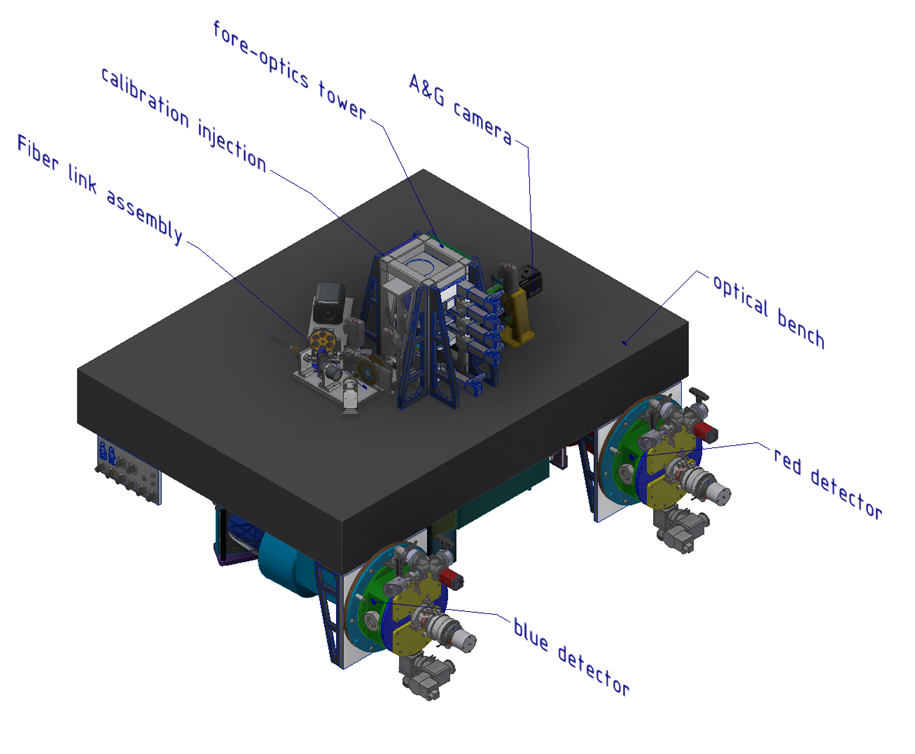}} \\
\center{ \resizebox{7.4cm}{!}{\includegraphics[clip=true]{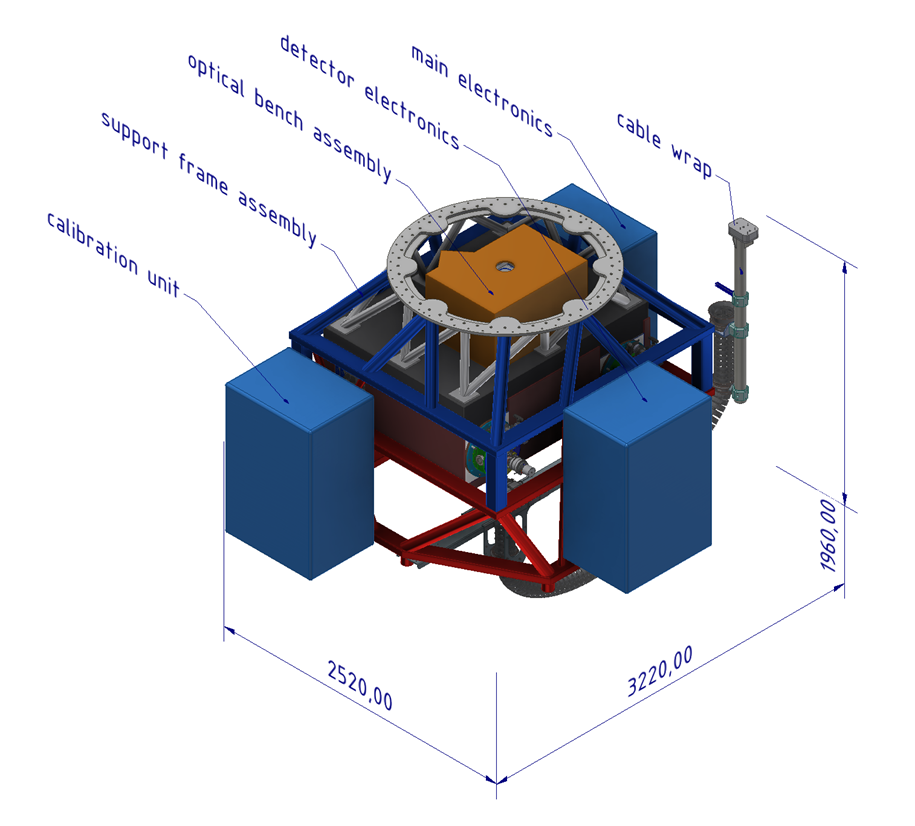}}}
\caption{\footnotesize
{\bf Mechanical concept for CUBES: The general layout of the main mechanical components is shown. For reference, the Telescope Adapter has a diameter of about 1.5 m. The Fore-Optics Assembly is located near the center of the bench, while the spectrograph opto-mechanics are ‘hanging’ on the bottom side of the same bench.}}
\label{fig:cubesmech}
\end{figure*}

In the current design, the CUBES main mechanical structure is divided into three main components: 1.  A telescope adapter, that provides a stiff connection between the Cassegrain telescope flange and the optical bench assembly; 2.  An optical bench that provides a stable platform for the spectrograph optics as well as for the foreoptics; 3.  An assembly to provide support for auxiliary  equipment  such  as  electronic racks, the calibration unit and vacuum equipment. In Fig.\,\ref{fig:cubesmech} the mechanical concept of CUBES is shown. 

\subsection{Software}
\label{sec:Software}
The CUBES instrument is designed including a  {\it ``software ecosystem''}, whose individual packages cooperate to support the users from the proposal preparation to the science-grade spectra: i) the Exposure Time Calculator (ETC), a web application used to predict the CUBES
exposure time required to achieve a given SNR;
ii) the Observation Preparation Software (OPS): a list of tools aimed to help the users identify the best instrument settings to achieve a scientific goal;
iii) the Instrument Control Software (ICS) and the Detector Control Software (DCS): devoted to the instrument devices and detectors control;
iv) the Data Reduction Software (DRS): a collection of recipes aimed to remove instrument signature from the science exposures, and produce the final calibrated 2D and 1D spectra;
v) the End-to-end Simulator (E2E): a package able to simulate realistic science exposures of a given science target, by taking into account the CUBES instrumental effects, and allowing early testing of the data reduction pipelines, as well as early validation of design decisions.

The above mentioned packages are developed according to the recently published ELT software standards, and will be based on the ELT Instrument Framework libraries and tools.

\section{Project and Management}
\label{sec:management} 
The CUBES consortium is composed of institutes from five countries:
\begin{itemize}
\item INAF - Istituto Nazionale di Astrofisica, Italy, (consortium leader)
\item IAG-USP - Instituto de Astronomia, Geofísica e Ciências Atmosféricas (primary Brazil partner) and LNA - Laboratório Nacional de Astrofísica (secondary Brazil partner), Brazil
\item LSW - Landessternwarte, Zentrum für Astronomie der Universtität Heidelberg, Germany
\item NCAC - Nicolaus Copernicus Astronomical Center of the Polish Academy of Sciences, Poland
\item STFC-UKATC - UK Astronomy Technology Centre, (primary UK partner) and Durham University Centre for Advanced Instrumentation (secondary UK partner), United Kingdom 
\end{itemize}
The Consortium organization is fairly standard, with a Principal Investigator (PI) with the ultimate responsibility for the project, acting as the formal contact point between ESO and the Consortium. The PI represents the leading technical institute, INAF. Each country is then represented in the managerial structure by one Co-PI; together they form the CUBES Executive Board (EB). The managerial aspects are delegated by the EB to the Project Manager (PM). The scientific aspects are delegated by the EB to the Project Scientist (PS). The project manager is supported by a System Engineer (SE) and by a Software System Engineer (SSE) who are in charge to supervise the overall system design. The SE and SSE work in close contact with the Instrument Scientist (IS) who makes sure that the adopted technical solutions match the foreseen instrument scientific needs. 

CUBES follows the standard project phasing for ESO instruments which is based on the stage-gate paradigm. Important decision points are project milestones (gates of the project) which mark the transition into a new stage when successfully completed. In the current plan CUBES will be available to the ESO user community in 2028.

\subsection{Public Engagement}
CUBES is an ambitious research program, and some of the scientific topics are related to the hottest open questions in modern astrophysics. Considering the vast discovery potential of the project, and the remarkable research and development technological activities, we consider good public communication as particularly important. Dissemination of science and technology is a fundamental part of our project. We have prepared a web page that is also a useful tool for the project as a whole\footnote{\url{https://cubes.inaf.it/home}}, and profiles in the main social media, i.e. Facebook, Twitter etc.

\section{Conclusions}
\label{sec:conclusions}  
The Cassegrain U-Band Efficient Spectrograph (CUBES) for the ESO VLT has been presented. Analysis of the design shows that it will deliver outstanding ($>40$\%) throughput across its bandpass, at a mean $R > 20$\,K (HR mode) and $R \sim 7K$ (LR mode). A fiber link from CUBES to UVES is being studied, which would provide the capability of simultaneous high-resolution spectroscopy at $\lambda > 420$\,nm. The CUBES design is able to address a large variety of scientific cases, from Solar System science to Cosmology, with no obvious technical showstopper. With contributions from institutes in five countries, the CUBES design is well placed to become the most efficient ground-based spectrograph at near-UV wavelengths, with science operations anticipated for 2028, opening a unique discovery space for the VLT for years to come.

\noindent {\bf{Affiliations}}\par
$^{6}$Ioffe Institute, HSE University, Saint-Petersburg, Russia\par
$^{7}$Universidade de São Paulo, IAG, Rua do Matão 1226, São Paulo, Brazil\par
$^{8}$Donostia International Physics Center (DIPC), Guipuzkoa, Spain\par
$^{9}$IKERBASQUE Basque Foundation for Science, Bilbao, Spain\par
$^{10}$University of Hull, E.A. Milne Centre for Astrophysics, Hull, UK\par
$^{11}$STFC - United Kingdom Astronomy Technology Centre (UK ATC), Edinburgh, UK\par
$^{12}$European Southern Observatory (ESO), ESO Headquarters
Karl-Schwarzschild-Str. 2 85748 Garching bei M\"unchen
Germany\par
$^{13}$Durham University, Department of Physics, Centre for Advanced Instrumentation, Durham, UK\par
$^{14}$INAF - Osservatorio Astronomico di Padova, Vicolo dell'Osservatorio 3, Padova, Italy\par
$^{15}$INAF - Osservatorio Astronomico di Roma, Via Frascati 33, Monte Porzio Catone, Italy\par
$^{16}$Centre for Astrophysics, University of Southern Queensland, Toowoomba 4350, Australia\par
$^{17}$INAF - Osservatorio Astronomico di Abruzzo, Via M. Maggini, I-64100 Teramo, Italy\par
$^{18}$INFN - Sezione di Pisa, Largo Pontecorvo 3, I-56127 Pisa, Italy\par
$^{19}$LNA/MCTI, Laboratorio Nacional De Astrofisica, Itajubá, Brazil\par
$^{20}$Dipartimento di Fisica, Sezione di Astronomia, Università di Trieste, Italy\par
$^{21}$Landessternwarte, Zentrum für Astronomie der Universität Heidelberg, Heidelberg, Germany\par
$^{22}$Centre for Extragalactic Astronomy, Durham University, Durham DH1 3LE, UK\par
$^{23}$Observat\'orio Nacional, Rua Gen. Jos\'e Cristino 77, S\~ao Crist\'ov\~ao, Rio de Janeiro, Brazil\par
$^{24}$Steward Observatory, University of Arizona, 950 N. Cherry Ave., Tucson, AZ, 85719\par
$^{25}$Nicolaus Copernicus Astronomical Center, Polish Academy of Sciences, Warsaw, Poland\par
$^{26}$Department of Astronomy, University of Geneva, Chemin Pegasi 51, Versoix, Switzerland\par
$^{27}$Consultant Astronomical Instrumentation, Alpenrosenstr.15, 85521 Ottobrunn, Germany\par
$^{28}$Italian Space Agency - Space Science Data Centre, via del Politecnico snc, Rome, Italy\par
$^{29}$Australian Astronomical Optics, Macquarie University, North Ryde, NSW 2113, Australia\par
$^{30}$European Space Agency (ESA), ESA Office, Space Telescope Science Institute, 3700 San Martin Drive, Baltimore, MD 21218, USA\par
$^{31}$Astrophysics Research Centre, Queen's University, Belfast, UK\par
$^{32}$University of Warwick, Department of Physics, Gibbet Hill Road, Coventry, CV7 4AL, UK\par
$^{33}$Goethe University Frankfurt, Institute for Applied Physics, Frankfurt am Main, Germany\par
$^{34}$Dipartimento di Fisica e Astronomia dell'Università, Vicolo dell'Osservatorio 3, Padova, Italy\par
$^{35}$Observat\'orio do Valongo, Universidade Federal do Rio de Janeiro, Brazil\par
$^{36}$INAF - Institute of Space Astrophysics and Planetology, Roma, Italy\par
$^{37}$Franco-Chilean Laboratory for Astronomy, Las Condes, Santiago, Chile\par
$^{38}$Institut d’Astrophysique de Paris, CNRS-SU, UMR 7095, 98bis bd Arago, Paris, France\par
$^{39}$Institute for Astronomy, University of Edinburgh, Royal Observatory, Edinburgh, UK\par
$^{40}$INAF - Osservatorio Astrofisico di Arcetri, Largo E. Fermi 5, 50125, Firenze, Italy\par
$^{41}$INAF - Osservatorio di Astrofisica e Scienza dello Spazio, Bologna, Italy\\

\begin{acknowledgements}
R.S. acknowledges support by the Polish National Science Centre through project 2018/31/B/ST9/01469.
We gratefully acknowledge support from the German Federal Ministry of Education and Research (BMBF) through project 05A20VHA.
B.B. acknowledges the FAPESP grant 2014/18100-4.
For the purpose of open access, the author has applied a Creative Commons Attribution (CC BY) licence to any Author Accepted Manuscript version arising from this submission.
\end{acknowledgements}

\bibliographystyle{aa}
\bibliography{mem_CUBES_bib}

\end{document}